\documentstyle[11pt,aaspp4]{article}

\renewcommand{\ga}{\,\raisebox{-0.4ex}{$\stackrel{>}{\scriptstyle\sim}$}\,}
\renewcommand{\la}{\,\raisebox{-0.4ex}{$\stackrel{<}{\scriptstyle\sim}$}\,}

\def\tinycaps#1{{\rm\scriptscriptstyle#1}}
\def\S{{\tinycaps S}}

\begin{document}
\title{LOW-MASS NORMAL-MATTER ATMOSPHERES OF STRANGE STARS
AND THEIR RADIATION
}
\author{Vladimir V. Usov}

\centerline{Department of Condensed Matter Physics, Weizmann Institute,
Rehovot 76100, Israel
} 
\medskip
\centerline{\it Accepted for publication in The Astrophysical Journal
Letters}

\begin{abstract}
The quark surface of a strange star has a very low emissivity
for X-ray photons. I find that a small amount of normal matter
at the quark surface with temperature in the range $10^7\la
T_{_{\rm S}} \ll mc^2/k\simeq 6\times 10^9$ K is enough
to produce X-rays with high luminosity, $L_X\simeq 10^{32}-
10^{34}(\Delta M/10^{-22}M_\odot)^2$ erg s$^{-1}$. For the total
atmosphere mass $\Delta M\sim (10^{-20}-10^{-19})M_\odot$, this
luminosity may be as high as the Eddington limit.
The mean energy of X-ray photons which are radiated from
such a low-mass atmosphere of a strange star
is $\sim 10^2(T_{_{\rm S}}/10^8\,{\rm K})^{0.45}\simeq 30-300$
times larger than the mean energy of X-ray photons which are
radiated from the surface of both a neutron star and a strange
star with a massive normal-matter envelope, $\Delta M\sim
10^{-5}M_\odot$, for a fixed temperature at the stellar core.
This raises the possibility that some black hole candidates
with hard X-ray spectra are, in fact, such strange stars with
a low-mass atmosphere. The X-ray emission from single strange
stars is estimated.

\keywords{stars: atmospheres - X-rays: stars -
radiation mechanisms: thermal}
\end{abstract}

\section{INTRODUCTION}
Strange stars have been proposed by Witten (\cite{W84}) as a new
class of astronomical compact objects. If Witten's idea is true,
at least some part of the compact
objects known to astronomers as pulsars, powerful accreting X-ray
sources, X-ray and $\gamma$-ray bursters, etc. might be strange
stars, not neutron stars as is usually assumed (Alcock et al.
\cite{AFO86}; Glendenning \cite{G90}; Caldwell \& Friedman \cite{CF91};
Madsen \& Olesen \cite{MO91}; Weber et al. \cite{WSWG96}).
Strange quark matter with the density of
$\sim 4\times 10^{14}$ g cm$^{-3}$ can exist, by hypothesis,
up to the surface of strange stars. The quark surface is a very poor
radiator at energies $\varepsilon_\gamma < 20$ MeV (Alcock et al.
\cite{AFO86}). But, the presence of "normal" matter (ions and electrons)
at the quark surface may restore the ability of the surface to radiate
soft photons (this is like painting with black paint on a silver surface).
There is an upper limit to the amount of normal matter at the quark surface,
$\Delta M\la 10^{-5}M_\odot$ (e.g., Glendenning \& Weber \cite{GW92}),
set by the requirement that the density of the inner layer
of normal matter at the quark surface
cannot be more than $\rho_n\simeq 4.3\times 10^{11}$ g cm$^{-3}$. Such
a massive envelope of normal matter with $\Delta M\sim 10^{-5}M_\odot$
completely obscures the quark surface.

If a neutron star is not too young (age $t\ga 10^2$ yr),
the stellar interior may be divided into two regions: the isothermal
core with density $\rho >\rho_e\sim 10^{10}- 10^{11}$ g cm$^{-3}$ and
the outer envelope with $\rho<\rho_e$
(e.g., Gudmundsson, Pethick \& Epstein
\cite{GPE1983}; Nomoto \& Tsuruta \cite{NT1987}; Schaaf \cite{S1990}).
At the core temperature $T_c\sim 10^7-10^9$ K, the temperature decreases
by a factor of $\sim 30-300$ in the envelope, from $T_c$
at the inner boundary of the envelope to $\sim 10^6\times
(T_c/10^8\,{\rm K})^{0.55}$ K at the neutron star surface.
Since $\rho_n\sim\rho_e$, for strange stars with massive envelopes,
$\Delta M\sim 10^{-5}M_\odot$, the temperature variation between the
quark surface and the surface of the normal-matter envelope
is more or less the same as
the core-to-surface temperature variation of neutron stars for a fixed
temperature at the stellar center. Besides,
if the strange quark matter is superfluid, the cooling
behavior of the quark core of strange stars is more or less similar to the
cooling behavior of the isothermal core of neutron stars (e.g.,
Weber et al. \cite{WSWG96}). Therefore, at an age $t\ga 10^2$ yr, photon
radiation from the surface of compact objects is not a good observational
signature of strange stars with massive envelopes of normal matter.

It was pointed out (e.g., Haensel, Paczy\'nski \&
Amsterdamski \cite{HPA}; Burrows \& Hayes
\cite{BH95}; Hartmann \& Woosley \cite{HW95}; Cheng \& Dai \cite{CD96})
that the temperature in the
interior of both neutron stars and strange stars at the moment of their
formation is very high, $T_c\sim$ a few $\times 10^{11}$ K. The mass of
gas which is ejected from the surface of such a hot neutron star
during $\sim 10$ s since its formation is $\sim 10^{-3}-
10^{-2}M_\odot$ (e.g., Woosley \& Baron \cite{WB92};
Levinson \& Eichler \cite{LE93}; Woosley \cite{W93}).
This value is considerably larger than the upper limit on the
mass of normal-matter envelopes of strange stars. The input physics for
calculations of gas outflow from the envelopes is similar for both hot
neutron stars and hot strange stars. Therefore, it is natural
to expect that if any normal matter remains
at the surface of a strange star at $t\gg 10$
s, its mass is many orders smaller than the maximum. In the process of
gas accretion onto such a strange star, the stellar atmosphere
has to pass through the stage with $\Delta M\ll 10^{-5}M_\odot$
before reaching the maximum,
$\Delta M\sim 10^{-5}M_\odot$, which was usually assumed in all studies
of the thermal structure and photon radiation of strange stars.
Below, photon radiation of a non-magnetic
strange star with a low-mass atmosphere,
$\Delta M\ll 10^{-5}M_\odot$, is considered.

\section{STRUCTURE AND PHOTON RADIATION OF THE ATMOSPHERE}

We consider here a star consisting of a core of strange quark matter
surrounded by a low-density atmosphere of normal matter with
the mass $\Delta M$, which is many orders of magnitude smaller than
$10^{-5}M_\odot$. The core acts on the atmosphere as a
heat reservoir. The thermal structure and photon radiation of the
atmosphere can be found by solving the heat transfer problem with
$T=T_{_{\rm S}}$ as a boundary condition at the inner layer of
normal matter. We assume that the temperature of the quark surface is
$T_{_{\rm S}}\ga 10^7$ K, and the hot gas of the atmosphere
emits mainly due to free-free transitions
(Gaetz and Salpeter \cite{GS83}).

A young strange star cools very rapidly due to intense neutrino
emission (e.g., Weber et al. \cite{WSWG96}). Therefore, we restrict
our consideration to the case of non-relativistic temperatures,
$T_{_{\rm S}}\ll T_0=mc^2/k\simeq 6\times 10^9$ K.
At such temperatures, the energy loss per unit gas volume by
bremsstrahlung radiation is

\begin{equation}
Q_{ff} = C_1N_iN_eZ^2T^{1/2}\,\,\,\,\,
{\rm erg\, s}^{-1}\,{\rm cm}^{-3} \ ,
\label{Q}
\end{equation}

\noindent where $N_i = \rho/m_pA$ is the ion density in cm$^{-3}$ ,
$N_e = N_iZ$ is the electron density in cm$^{-3}$, $m_p$ is the
proton mass, $A$ is the mass number of ions, $Z$ is their
electrical charge, $T$ is the temperature in K, $\,C_1 \simeq
1.4\times 10^{-27} g_o(T)$ and $g_o(T)$ is  the frequency
averaged Gaunt factor, which is
in the range 1.1 to 1.5. Choosing a value of 1.2 for $g_o(T)$
will give an accuracy to within about 20\% (Karzas and Latter
\cite{KL61}).

At $T\ll T_0$, the scale height, $\Delta x$,
of the atmosphere is very small compared to
the stellar radius $R$, and a plane-parallel approximation may be
used. In this approximation all parameters depend on only one
coordinate, $x$, which is the distance from the quark surface.

The set of equations which describes the structure of the atmosphere
will be the equation of hydrostatic equilibrium
and the energy transport equation:

\begin{equation}
{dP\over dx}=-{\rho GM\over R^2}\,,\,\,\,\,\,\,\,\,\,\,
{dF\over dx}=- Q_{ff}\,,
\label{dP}
\end{equation}

\noindent where $P= (N_e + N_i)kT= {\rho kT /m_p\mu}$ is the gas
pressure, $\mu$=$A/(1+Z)$ is the mean molecular weight,
$G$ is the gravitational
constant, $M$ is the stellar mass, and $F$ is the heat flux due to both
thermal conductivity and convection (Schwarzschild \cite{S58}).
The absorption of radiation is ignored in the energy
transport equation because in our case the atmosphere is
optically thin for the bulk of radiation (see below).

\subsection{Isothermal atmosphere}

The characteristic time of heat conduction in the
atmosphere is

\begin{equation}
t_{\rm heat}\simeq {k(1+Z)N_e(\Delta x)^2\over Z\eta}\,\,\,\,
{\rm s}\,,
\label{tauh}
\end{equation}

\noindent
and the characteristic cooling time of the hot
atmospheric plasma via bremsstrahlung is

\begin{equation}
t_{\rm cool}\simeq 1.5\times 10^{11}{(1+Z)T^{1/2}\over Z^2N_e}\,
\,\,\,{\rm s}\,,
\label{tauc}
\end{equation}

\noindent where $\eta\simeq 10^{-6}Z^{-1}T^{5/2}$ erg s$^{-1}$
cm$^{-1}$ K$^{-1}$ is the coefficient of heat conductivity
for a rarefied  totally-ionized plasma (Spitzer \cite{S67}).

When $t_{\rm heat}\ll t_{\rm cool}$, the atmosphere is nearly
isothermal, $T\simeq T_{_{\rm S}}$, and the equation of hydrostatic
equilibrium (\ref{dP}) can be integrated immediately:

\begin{equation}
\rho=\rho_0\exp \left(-{x\over \Delta x}\right)\,,\,\,\,\,{\rm where}
\,\,\,\,\Delta x={R^2kT_{_{\rm S}}\over GM\mu m_p}\,.
\label{Dx}
\end{equation}

In this case the photon luminosity of the atmosphere is

\begin{equation}
L=4\pi R^2\int_{0}^{\infty}Q_{ff}dx\simeq  {4\times 10^{33}Z^3
\over A(1+Z)}
\left({R\over 10^6\,{\rm cm}}\right)^{-4}\left({M\over M_\odot}\right)
\left({T_{_{\rm S}}\over 10^8\,{\rm K}}\right)^{-1/2}
\left({\Delta M\over 10^{12}\,{\rm g}}\right)^2\,\,\,\,{\rm erg\,s}^
{-1}\,,
\label{L2}
\end{equation}

\noindent
where $\Delta M =4\pi R^2\rho_0\Delta x$ is the total mass of the
atmosphere.

Using equations (\ref{tauh})
and (\ref{tauc}), the condition $t_{\rm heat}\ll t_{\rm cool}$ may be
written as a limitation of the atmosphere mass:
$\Delta M\ll\Delta M_1$, where

\begin{equation}
\Delta M_1\simeq 7\times 10^{11}{A\over Z^{2}}
\left({T_{_{\rm S}}\over 10^8\,
{\rm K}}\right)^{3/2}\left({R\over 10^6\,{\rm cm}}\right)^2\,\,\,
{\rm g}\,.
\label{DM}
\end{equation}

\subsection{Convective atmosphere}

The heat flux due to thermal conductivity,
which is responsible for the heat transport in the atmosphere
when $\Delta M< \Delta M_1$, is $F= - \eta dT/ dx$.
Using this, we can rewrite the equation
of hydrostatic equilibrium (\ref{dP}) in the form:

\begin{equation}
{d\rho\over dx}=-{\rho\over T} \left({GMm_p\mu\over R^2k}-{F\over
\eta}\right)\,.
\label{drho}
\end{equation}

In steady state, which we assume in this paper,
we have the following boundary condition for $F$ at $x=0$:
$F\vert_{x=0}={L/ 4\pi R^2}$. From this condition and
equation (\ref{drho}), the gradient of the
density at the quark surface, $x=0$, is

\begin{equation}
{d\rho\over dx}\vert_{x=0}=-{\rho_0\over T_{_{\S}}R^2}
\left({GMm_p\mu\over k}-
{L\over 4\pi\eta}\right)\,.
\label{drho0}
\end{equation}

If the heat transport in the atmosphere
is only due to thermal
conductivity and the photon luminosity of the atmosphere is higher than

\begin{equation}
L_1={4\pi GMm_p\mu \over k}\eta\vert_{T=T_{_{\rm S}}}\simeq 3\times
10^{33}{\mu\over Z}\left({M\over M_\odot}\right)\left({T_{_{\rm S}}
\over 10^8\,{\rm K}}\right)^{5/2}\,\,\,\,{\rm erg\,s}^{-1}\,,
\label{L-1}
\end{equation}

\noindent
then the density of normal matter near the quark surface should
increase with the distance from the surface, $(d\rho/dx)\vert_{x=0}>0$.
However, such a density behaviour is unstable, and convection
develops in the atmosphere at $L>L_1$. The value of $L_1$
coincides with the photon luminosity (\ref{L2}) after
substitution $\Delta M=\Delta M_1$ from equation
(\ref{DM}) into equation (\ref{L2}).

If convection takes place in the stellar atmosphere, it is a good
approximation, as a rule, to say that the temperature gradient is
equal to the adiabatic one, i.e. $PT^{\gamma/(1-\gamma)}=$ constant,
where $\gamma$ is the ratio of the specific heats at constant pressure
and  at constant volume. In this case, from equation (\ref{dP}),
the temperature and density distributions are

\begin{equation}
T=T_{_{\rm S}}\left[{1-{(\gamma -1)x\over\gamma\Delta x}}
\right], \,\,\,\rho=\rho_0\left({T\over T_{_{\rm S}}}\right)^
{1/(\gamma -1)},
\label{tem}
\end{equation}

\noindent where $\Delta x$ is determined by equation (\ref{Dx}).

Using equations (\ref{Q}) and (\ref{tem}), one can get
the photon luminosity of the convective atmosphere
$\tilde L={4\gamma L/( 3\gamma + 1)}$,
where $L$ is the photon luminosity of the isothermal
atmosphere which is given by equation (\ref{L2}). Since
$\gamma \geq 1$, the value of $\tilde L$ is in the range $L\leq \tilde L<
{4\over 3}L$. For a rarefied  totally-ionized plasma we have
$\gamma = {5\over 3}$ and $\tilde L=
{10\over 9}L$. As noted earlier, the accuracy of equation
(\ref{Q}) for the energy loss $Q_{ff}$ is about 20\%. Moreover,
we did not take into account the general relativity effects
which are $\sim 20$\%. Therefore, the accuracy of our
calculations of bremsstrahlung radiation from the
atmospheres of strange stars is several times ten of percent.
We can see that the difference between $L$ and $\tilde L$
is within the accuracy of our calculations.

The adiabatic approximation may be used
to estimate the photon luminosity of the convective atmosphere
only if the characteristic time of
convection $t_{\rm conv}\simeq \Delta x/v_{\rm conv}$, which is the
main process of heat transport at $\Delta M>\Delta M_1$, is smaller than
the characteristic cooling time $t_{\rm cool}$ for the atmospheric
plasma, where $v_{\rm conv}$ is
the convective velocity which is limited by the velocity of sound,
$v_{\rm conv} \la c_s=(\gamma P/\rho )^{1/2}$. Using this and equation
(\ref{tauc}), the condition $t_{\rm conv}<t_{\rm cool}$ may be
written in the form: $\Delta M<\Delta M_2$, where

\begin{equation}
\Delta M_2\simeq {4\times 10^{12}A\over Z^2\mu^{1/2}}
\left({T_{_{\rm S}}\over 10^8\,
{\rm K}}\right)\left({R\over 10^6\,{\rm cm}}\right)^2\,\,\,\,
{\rm g}\,.
\label{DM2}
\end{equation}

At $\Delta M=\Delta M_2$, the photon luminosity is

\begin{equation}
L_2\simeq 6\times 10^{34}{(1+Z)^2\over Z^3}\left({M\over M_\odot}
\right)\left({T_{_{\rm S}}\over 10^8\,{\rm K}}\right)^{3/2}\,\,\,\,
{\rm erg\,s}^{-1}\,.
\label{LDM2}
\end{equation}

For $M\simeq 1.4 M_\odot$, $T_{_{\rm S}}\simeq 2\times 10^8$ K and
$Z=1$ the value $L_2$ is $\sim 10^{36}$ erg s$^{-1}$ that is only
two orders of magnitude smaller than the Eddington limit
$L_{\rm Edd}\simeq 1.3\times 10^{38} (A/ Z)({M/ M_
\odot})$ erg s$^{-1}$.

The mean free-free optical depth of the atmosphere is

\begin{equation}
\tau_0\simeq \alpha_{ff} \Delta x
\sim 10^{-9}
\left({L\over 10^{34}\,{\rm erg\,s}^{-1}}\right)
\left({T_{_{\rm S}}\over 10^8\,{\rm K}}\right)^{-4},
\label{tau}
\end{equation}

\noindent where $\alpha_{ff}\simeq 10^2T_{_{\rm S}}^{-4}Q_{ff}$ is
the Rosseland mean of the free-free absorption coefficient.
At $T_{_{\rm S}} >$ a few $\times 10^7$ K, the atmosphere
is optically thin, $\tau_0 \ll 1$, up to $L\sim L_{\rm Edd}$.

For $\Delta M>\Delta M_2$, both thermal conductivity and
convection are not able to account for the cooling of
atmospheric matter, and
a thermal instability develops in the atmosphere. As a
result of this, the atmosphere cannot be in hydrostatic
equilibrium during a time larger than $\sim t_{\rm cool}$,
and it is strongly variable on a timescale of
a few $\times (2R^2\Delta x/GM)^{1/2}$ $\sim
10^{-4}(T_{_{\rm S}}/10^8\,{\rm K})^{1/2}$ s.
Consideration of this variability and
estimation of the photon luminosity
at $\Delta M > \Delta M_2$ are under way
and will be published elsewhere. Here, it is worth noting only that
at $T_{_{\rm S}} >$ a few $\times 10^7$ K and $\Delta M>\Delta M_2$
the tendency of the photon luminosity to increase with
increase of $\Delta M$ has to be held up to $L =L_{\rm Edd}$.

\section{CONCLUSIONS AND DISCUSSION}

The photon luminosity of a strange star with a low-mass
atmosphere, $\Delta M<\Delta M_2$, is given by
equation (\ref{L2}) with the accuracy of several times ten of
percent irrespective of the atmosphere structure.
The photon luminosity may be very high, up to $\sim  L_{\rm Edd}$.
It is very important for discovery of strange stars that the mean
energy of X-ray photons which are radiated from such stars is
substantially larger than the mean energy of X-ray photons which
are radiated from the surface of both a neutron star and a strange
star with a massive envelope of normal matter, $\Delta M\sim
10^{-5}M_\odot$, for a fixed temperature at the stellar core.

A source of X-rays may be a strange star
if it meets the following criteria:

1. The X-ray emission (or at least one of its components) may
be fitted by thermal emission of optically-thin plasma
at $kT$ up to $\sim 10^2$ keV.

2. The X-ray flux is variable at the X-ray luminosity $L_X> L_2
\simeq (0.05-5)\times 10^{35}(T/10^8\,{\rm K})^
{3/2}$ erg s$^{-1}$ depending on the composition of the atmosphere.
This variability is either irregular or quasi-periodic.

3. The mass of the compact X-ray source is on the high side
for neutron star masses because
conversion of a neutron star to a strange star requires a very high
density at the center of the neutron star (e.g., Alcock et al.
\cite{AFO86}).

A few enigmatic X-ray sources which are considered as
black hole candidates (e.g., Cherepashchuk \cite{C96})
answer these criteria and may be, in fact, strange stars with a
low-mass atmosphere. They are 1E 1740.7 - 2942; GRS 1915 - 105,
GRO J0422 + 32, GX 339 - 4 and SS 433. Some other
powerful X-ray sources which are black hole candidates as well,
for example Cyg X-1, have both
hard X-ray spectra which may be fitted by emission of
optically-thin plasma and strong X-ray flux variability.
The existent lower limits to the mass of these
X-ray sources are substantially higher than the Oppenheimer-Volkoff
limit for a strange star (Cherepashchuk \cite{C96}). However, these
objects may be, in principle, a triple system with strange
star (cf. Bahcall et al. \cite{B1974}).

The thermal energy of a strange star itself is not enough for the star
to be a powerful X-ray source, $L_X\sim (0.1-1)L_{\rm Edd}$,
for a long time, $t\ga 10^4-10^5$ yr, and accretion of gas
onto the strange star is necessary to account for
such a strong prolonged X-ray emission. The kinetic energy
of the accreted gas may be transformed into emission of the
strange star atmosphere in the following way. Let the magnetic
field at the stellar surface be $\sim (0.3-1)\times
10^{11}$ Gauss. This field canalizes the gas motion along the field
lines to the quark surface. In this case, the kinetic energy of ions
at the surface is about hundred MeV per nucleon that is $\sim 5$ times
more than the Coulomb barrier at
the quark surface (Alcock et al. \cite{AFO86}). Accreted particles
penetrate through the quark surface (cf. below), and they are
dissolved into quark matter. As a result, the quark core is heated
at the magnetic poles. The process of heat transport through the core
is very fast due to a very high heat conductivity of quark matter, and
therefore, the quark core is nearly isothermal. Then, the energy
which is released in the process of gas accretion is radiated from
the normal-matter atmosphere more or less isotropically just as
it is discussed above.

The total atmosphere mass which restores the ability of
the quark surface to radiate X-ray photons is extremely small. At first
sight, even if the rate of gas accretion onto a hot strange
star is very small, the atmosphere mass increases rapidly up the the
value when the photon luminosity is $\sim L_{\rm Edd}$.
But this does not necessarily happen. The point
is that ions of accreted gas in the process of their motion through
the atmosphere collide with the atmosphere ions and draw them
into the quark surface. As a result, a steady state
of the atmosphere may be achieved before the photon luminosity is
$\sim L_{\rm Edd}$. Let us estimate the photon luminosity at such a
steady state. For definiteness, we assume that a single strange
star with $M\simeq 1.4 M_\odot$ is
at rest in a uniform ionized gas of pure hydrogen. We shall
assume that the gas is at rest at infinity with the density of $\sim 1$
cm$^{-3}$ and the temperature of $\sim 10^4$ K. This situation will
correspond to accretion in a  typical interstellar H II region.
In this case, the accretion rate is $\dot M\simeq 2\times 10^{10}$
g s$^{-1}$ (e.g., Shapiro \& Teukolsky \cite{ST83}).
The efficiency for radiation of the gas during its
accretion is very low, i.e. the gas motion is nearly adiabatic.
The mean kinetic energy of accreted protons at the surface is $E_{\rm
kin}\simeq 100$ MeV. The temperature of accreted gas near the stellar
surface is $T_h\simeq 10^{11}$ K ($kT_h\simeq 10$ MeV). The bulk of
accreted protons passes through the quark surface. Protons
can be reflected from the quark surface back into
the atmosphere only if in the frame of the star the kinetic energy
of their radial motion to the stellar surface
is smaller than  the Coulomb barrier. Taking this into account and
assuming that the accreted protons are Maxwellian,
the rate of the atmosphere mass accumulation is $\dot M_a\simeq
\exp (- E_{\rm kin}/kT_h)\dot M\simeq \exp (-10)\dot M$.
The characteristic time of the atmosphere mass decrease due to
bombardment by the accreted protons is $\Delta t_b\simeq
(n_a \sigma v_a)^{-1}\simeq (4\pi R^2m_p/\dot M\sigma)$, where
$n_a\simeq \dot M/ (4\pi R^2v_am_p)$ is the density of accreted
protons at the surface, $v_a$ is their velocity and $\sigma \simeq
10^{-26}$ cm$^2$ is the cross section for proton-proton collision  at
energies of $\sim 100$ MeV. The atmosphere mass at the steady state
is $\Delta M_{\rm st}\simeq \dot M_a\Delta t_b\simeq 4\pi\exp
(- E_{\rm kin}/kT_h)R^2m_p\sigma ^{-1}\simeq 10^{12}$ g
which does not depend on the accretion rate. The
steady state may be reached in $\sim\Delta t_b\sim 10^5$ s.
From equation (\ref{L2}), for $T_{_{\rm S}}\sim 10^7-10^8$ K
the expected X-ray luminosity of a single strange star
is $L_X\sim 10^{34}$
erg s$^{-1}$, that is of the order of X-ray luminosity of a
single neutron star in $\sim 10^2-10^4$ yr after its formation.
But, in the case of such a strange star
the expected X-ray spectrum is much harder than
the X-ray spectrum of the thermal radiation from a single
neutron star with the same luminosity.

\begin{acknowledgements}
I thank R. McCray, M. Milgrom and M. Rees for many helpful discussions.
\end{acknowledgements}

\end{document}